\documentclass[]{spie}  

\usepackage{amsmath,amsfonts,amssymb}
\usepackage{graphicx}
\usepackage[] {subfigure}
\usepackage[colorlinks=true, allcolors=blue]{hyperref}
\usepackage{wasysym}
\usepackage{aas_macros}
\usepackage{orcidlink}
\title{The legacy of the IXPE instrument and prospects for the next generation of polarimetric photoelectric X-ray detectors}

\author[a,\orcidlink{0000-0001-9404-1601}]{Hemanth Manikantan}
\author[b,\orcidlink{0000-0003-3919-3068}]{Vladislavs Plesanovs}
\author[a,\orcidlink{0000-0002-7781-4104}]{Paolo Soffitta}
\author[c]{Dennis Sauerland}
\author[c]{Reinhard Beck}
\author[a,\orcidlink{0000-0003-4925-8523}]{Enrico Costa}
\author[a,\orcidlink{0000-0002-3013-6334}]{Ettore Del Monte}
\author[b,\orcidlink{0000-0001-5836-6118}]{Klaus Desch}
\author[a,\orcidlink{0000-0002-1159-4460}]{Alessandro Di Marco}
\author[a,\orcidlink{0000-0003-1533-0283}]{Sergio Fabiani}
\author[a,\orcidlink{0000-0003-1074-8605}]{Riccardo Ferrazzoli}
\author[b,\orcidlink{0000-0002-4767-1392}]{Markus Gruber}
\author[a,d,\orcidlink{0009-0001-1972-7774}]{Saba Imtiaz}
\author[b,\orcidlink{0000-0001-5979-6996}]{Jochen Kaminski}
\author[a]{Alessandro Lacerenza}
\author[a,\orcidlink{0000-0003-3331-3794}]{Fabio Muleri}
\author[a,e,\orcidlink{0000-0003-0411-4243}]{Ajay Ratheesh}
\author[a]{Alda Rubini}
\affil[a]{INAF Istituto di Astrofisica e Planetologia Spaziali, Via del Fosso del Cavaliere 100, Rome, Italy}
\affil[b]{Physikalisches Institut, Universität Bonn, Nussallee 12, Bonn, Germany}
\affil[c]{Helmholtz-Institut für Strahlen- und Kernphysik, Bonn, Germany}
\affil[d]{Padua University, Via Marzolo 8, Padua, Italy}
\affil[e]{Physical Research Laboratory, Thaltej, Ahmedabad, Gujarat 380009, India}

\authorinfo{Further author information: (Send correspondence to Paolo Soffitta)\\Paolo Soffitta: E-mail: paolo.soffitta@inaf.it}

\begin{document} 
\maketitle

\begin{abstract}
Imaging X-ray polarimetry with IXPE has demonstrated the scientific potential of the technique but also revealed the need for significant detector upgrades, particularly with the read-out ASIC that images photoelectron tracks and possibly the multiplication stage. Building on this experience, we are developing a next generation three-dimensional photoelectron track polarimeter based on the GridPix detector, originally developed for axion and axion-like-particle searches. We report on the current status of prototype development and preparations for the ion-irradiation tests. Preliminary proton beam irradiation runs at the Bonn Isochronous Cyclotron facility of the University of Bonn verified both this generation of ASIC's tolerance to high radiation doses present in space and the capability of the cyclotron facility to operate at sufficiently low rates for controlled tests.
\end{abstract}


\section{INTRODUCTION}

The Imaging X-ray Polarimetry Explorer (IXPE) \cite{Weisskopf2022,Soffitta2021} has produced a wealth of new results, observing nearly $\sim $100 sources, most of which show significant polarized emission. IXPE has opened a new window in X-ray astronomy and, more broadly, astrophysics, enabling constraints on emission processes and the geometry of compact objects.

IXPE was conceived as an SMEX mission and designed under tight mass, power, and volume constraints to accommodate the initial request from NASA (PegasusXL). Consequently, each mirror module provided an effective area\cite{Ramsey2022} of only $\sim$200 cm$^{2}$ with thin, lightweight shells and a correspondingly limited count rate capability. However, when targeting bright X-ray sources, most notably Sco X--1 \cite{LaMonaca2024} and Swift J1727.8--1613 \cite{Veledina2023}, the use of an on-board flux-attenuation filter proved necessary.

The dominant bottleneck arises from the dead-time performance of the XPOL ASIC \cite{Bellazzini2006}, only partially mitigated by the evolved design presented in \cite{Minuti2023}. This constitutes a significant limitation in view of next-generation X-ray mirrors, which are expected to deliver a larger effective area $\sim $10-20 $\times$, and therefore demand substantial advances in the ASICs responsible for photoelectron track charge imaging.

Furthermore, a major issue of the IXPE detectors\cite{Baldini2021} results from the deposition of ions onto the dielectric (liquid crystal polymer) of the Gas Electron Multiplier (GEM\cite{Sauli1997,Tamagawa2006}), resulting in a decrease in the gas gain when observing bright sources ($\sim$ 0.5 crab or more). An on-board calibration system\cite{Ferrazzoli2020} allows one to mitigate this effect, albeit not completely.

We are investigating whether a different multiplication structure (InGrid\cite{Krieger2013}) above the ASIC solves this problem.    
With this design, a metallic mesh is supported by the microscopic pillars of SU-8 with a separation of 80 $\mu$m from the ASIC. This structure, in principle, prevents the build-up of charge on the metallic mesh, offering instead a $>$ 3-fold smaller diffusion with respect to the GEM, and a nearly perfect alignment of the mesh holes with the underlying pixelated CMOS ASIC. A protective layer of Si$_3$N$_4$ is usually deposited on the ASIC surface for protection against discharge. The effect of Si$_3$N$_4$ on the gain stability with flux needs to be assessed. GridPix\cite{vanderGraaf2007,Kaminski2012} is a gas detector that couples InGrid with Timepix3\cite{Pokela2014}. In the following section, we describe its main features and how they can be exploited for sensitive X-ray polarimetry.

\section{A dead-time free 3-D imaging X-ray photoelectric polarimeter}

The Medipix Collaboration devised the Timepix3 ASIC, a general-purpose CMOS pixel detector with a \(55\,\mu\mathrm{m}\) pitch. Each pixel integrates an ADC that provides the digitized charge via the time-over-threshold (ToT) measurement, while the pixel trigger time (time-of-arrival, ToA) is recorded simultaneously with a \(1.56\,\mathrm{ns}\) time resolution.

The per-pixel dead time is:
\begin{equation}
t_{\mathrm{dead}} \simeq 475~\mathrm{ns} + N_{\mathrm{ToT}}\times 25~\mathrm{ns},    
\end{equation}
where \(N_{\mathrm{ToT}}\) is the number of cycles of the \(40~\mathrm{MHz}\) clock during which the signal remains above a programmable threshold. 

In the data-driven (sparse) mode, both ToA and ToT, together with the pixel coordinates, are transmitted by the chip for every hit that crosses a preset threshold. The sparse readout is effectively dead-time free up to \(\sim 40~\mathrm{Mhits\,s^{-1}\,cm^{-2}}\). In this mode, the combined ToA and ToT information can be exploited to reconstruct the three-dimensional evolution of the track recorded by the ASIC after drift to the sensitive plane.

Table~\ref{tab:GPD_vs_GridPix} summarizes the performance of representative state-of-the-art ASICs for X-ray polarimetry based on photoelectron tracking in gas.

\begin{figure}[htpb]
\centering
\subfigure[\label{fig:3DTrack}]{\includegraphics[scale=0.8]{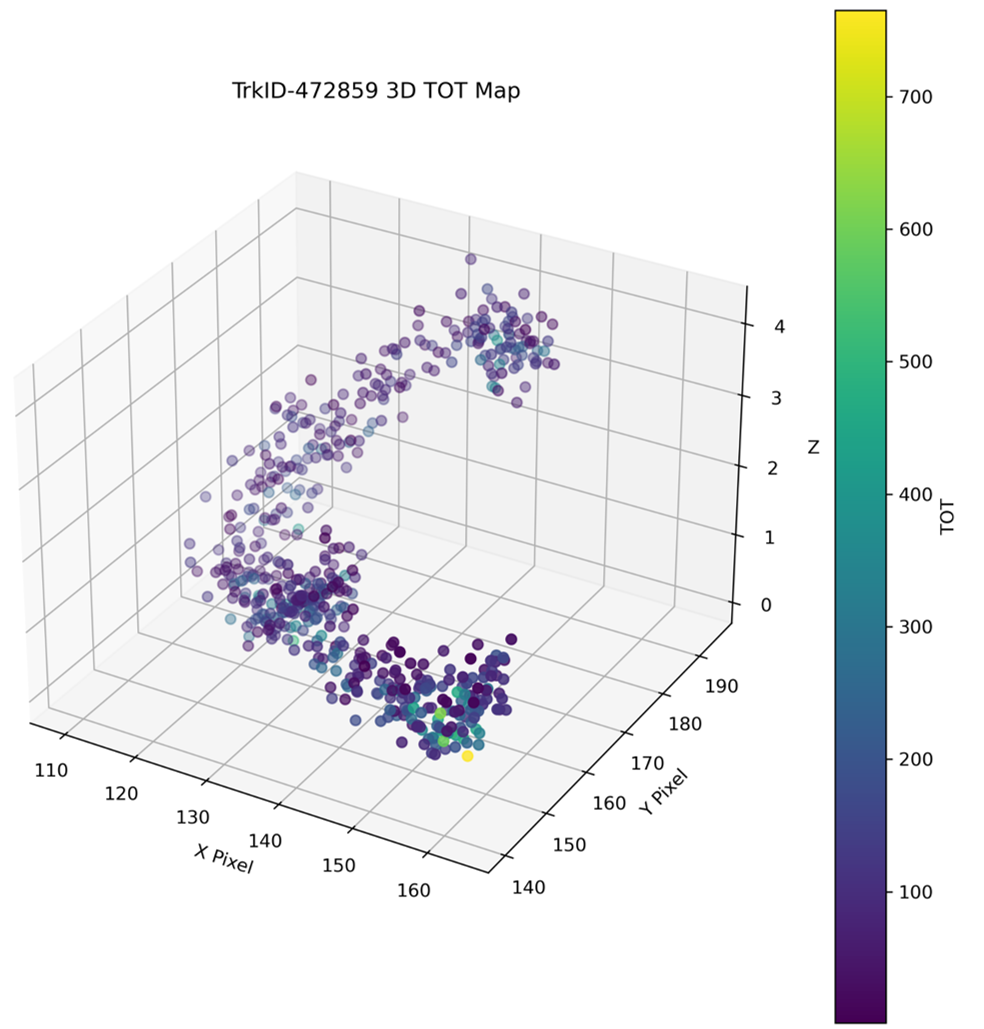}}
\hspace{1cm}
\subfigure[\label{fig:Rate}]{\includegraphics[scale=0.3]{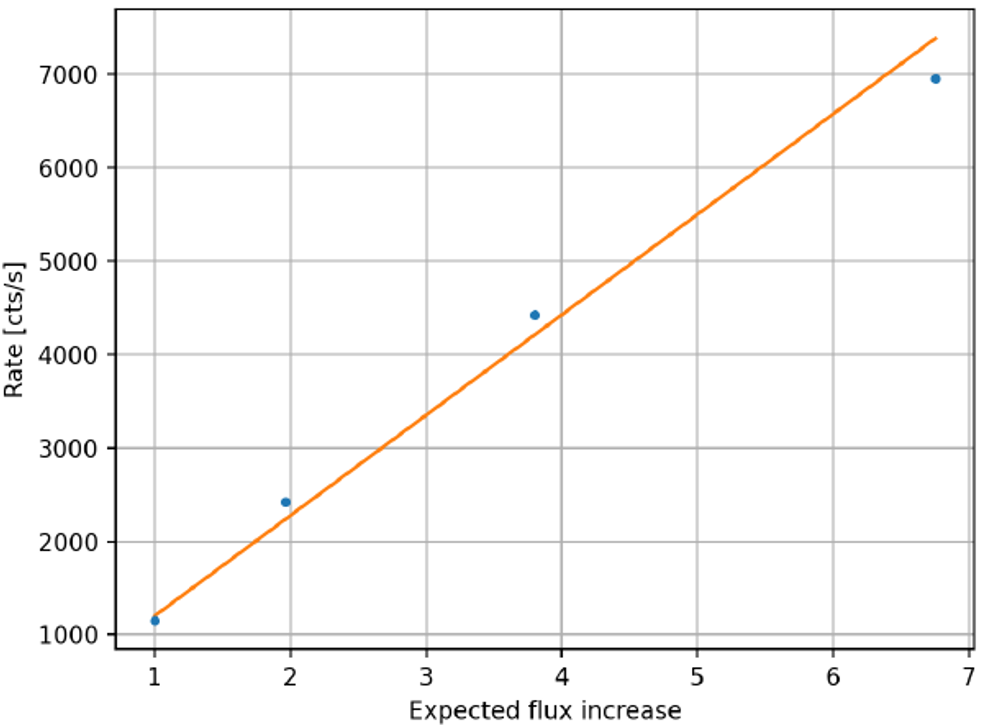}}
\caption{({\bf a}) A real 3-D photoelectron track imaged by GridPix from a 17.4 keV photon in an Ar-DME 80-20 mixture at 1 atm and 2 cm of drift length. The X, Y, and Z axes are shown in units of pixel pitch (55 $\mu$m), with the collected charge (ToT cycles) color coded. ({\bf b}) Measured counting rate as a function of the X-ray tube anodic current. The counting rate is linearly dependent on the anode current up to at least 7000 counts per second, setting a minimum 100-times improved rate capability with negligible dead-time with respect to the IXPE detector.} \label{fig:3dtrk_rate}
\end{figure}

\begin{table}[ht!]
\centering
\begin{tabular}{|c|c|c|} \hline 
& \textbf{GPD}& \textbf{GridPix}\\ \hline 
\textbf{Pattern}&  Hex 50 $\mu$m pitch& 55$\times$ 55 $\mu$m$^2$ pitch\\ \hline 
\textbf{Amplitude Signals Output}& Analog & Digital (ToT) \\\hline 
\textbf{Size}&  1.5 $\times$ 1.5 cm$^{2}$& 1.4 x 1.4 cm$^2$\\ \hline 
\textbf{Track imaging}&  2-D& 3-D\\ \hline 
\textbf{Maximum rate}  for 10 $\%$ dead time fraction&  100 (700)& 10$^6$ (100 pixels track)\\ \hline 
\textbf{Diffusion in Transfer gap} (DME, $\mu$m$_\mathrm{rms}$)&  55 & 8\\ \hline 
\textbf{Noise per pixel}  (e$_\mathrm{rms}$)&  50 (30)& 60\\ \hline 
\textbf{Threshold dispersion}  (e$_\mathrm{rms}$)& Non equaliz.& 35 (equaliz.)
\\ 
\hline 
\textbf{Minimum Threshold}  (e$^-$)& 2300 (150) & 500 (ToT/ToA)\\ 
\hline
\end{tabular}
\caption{Table of comparison\cite{Soffitta2024} of performances  between IXPE/GPD and GridPix (with Timepix3 ASIC\cite{Pokela2014}). In parentheses of the GPD column are the expected performances of XPOL III \cite{Minuti2023}, a new version of the ASIC on-board IXPE.}
\label{tab:GPD_vs_GridPix}
\end{table}
At INAF-IAPS, we already tested a GridPix detector provided by the University of Bonn. This detector, employed in axion searches, is currently used for the CAST\cite{Krieger2014} project and will be employed in the future for Baby-IAXO\cite{Altem2023}. We succeeded in our goal and demonstrated dead-time free acquisition above 7000 cts s$^{-1}$, which is a factor 100 larger than the count rate of Crab Nebula in one IXPE detector unit (see Fig. \ref{fig:Rate}). We also show a sample image of a real 3-D track as imaged by the GridPix during our test in Fig.~\ref{fig:3DTrack}.

\begin{figure}[htpb]
\centering
\subfigure[\label{fig:InGrid}]{\includegraphics[scale=0.3]{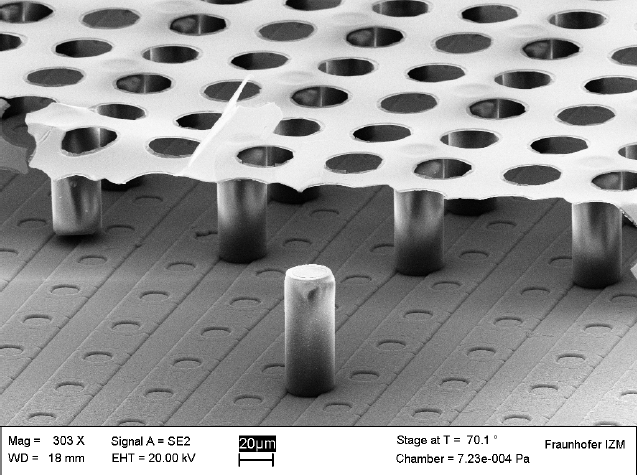}}
\hspace{1cm}
\subfigure[\label{fig:diff}]{\includegraphics[scale=1.0]{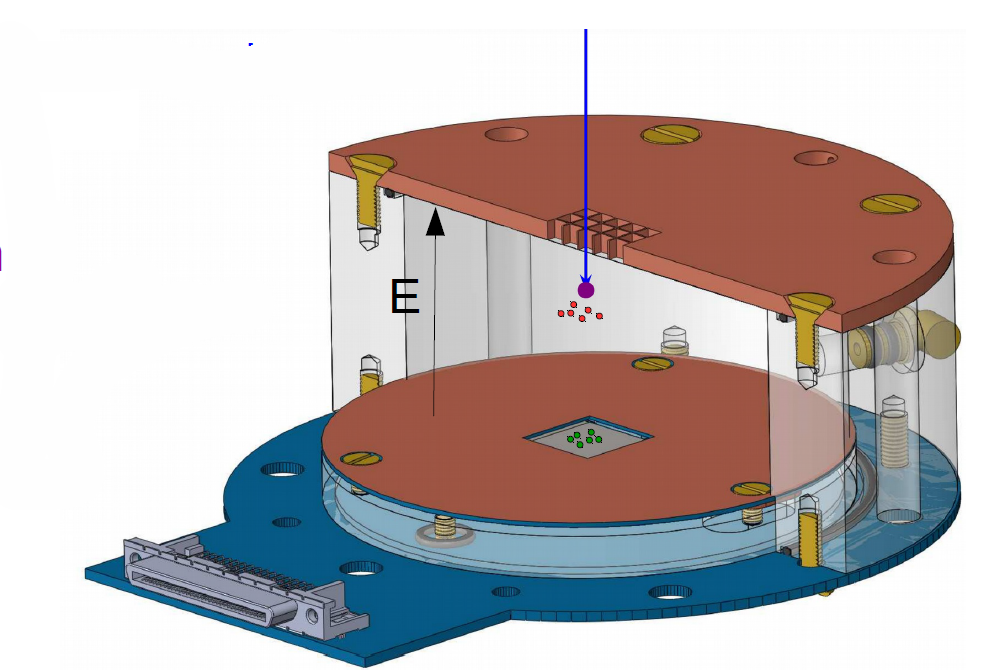}}
\caption{({\bf a}) The InGrid structure. The SU-8 pillars are 80 $\mu$m high, each hole is precisely on top of the ASIC pixel. ({\bf b}) A copper Anode surrounds the central square hole, which hosts the InGrid structure and Timepix3 ASIC, guaranteeing  uniformity of the Drift field at the edge of InGrid.} \label{GriPix&InGrid}
\end{figure}

Fig.~\ref{GriPix&InGrid} shows both the InGrid structure and the design of the detector body, as tested at INAF-IAPS laboratory. The InGrid structure was also mechanically tested with both random vibration tests at GEVS (General Environmental Verification Standard) levels and in the thermal vacuum chamber with 8 cycles between 4$^\circ$ and 40$^\circ$ and one at 100$^\circ$, the latter to check resilience for additional baking temperature for future sealed operations. 

The time-of-arrival (ToA) recorded for each pixel is converted into a spatial coordinate by using the expected electron drift velocity evaluated at the nominal drift field. The latter is obtained from the latest CERN-developed simulation toolchain—Garfield++\footnote{\url{https://garfieldpp.docs.cern.ch/}} together with MAGBOLTZ\footnote{\url{https://magboltz.web.cern.ch/magboltz/}}—to derive up-to-date electron-transport properties for the gas mixtures employed during the tests.

As already shown for one-dimensional gas TPCs proposed as astronomical X-ray polarimeters\cite{Black2022}, an experimental determination of the drift velocity will require additional dedicated measurements. In a one-dimensional gas TPC, owing to its peculiar geometry, this parameter directly affects the azimuthal modulation used to reconstruct the photoelectron emission direction; in the case of GridPix, by contrast, it primarily affects the reconstructed \(z\)-coordinate, which is typically orthogonal to the polarization-sensitive plane.

Part of the results from the first test run were already reported\cite{Ratheesh2024} without exploiting the full 3-D capabilities of the device.

\section{Motivation for an ion-beam irradiation} 
The most critical on-ground qualification for a detector’s survivability in orbit is its response to single-event effects (SEEs)—specifically single-event upset (SEU) and single-event latch-up (SEL)—together with its radiation tolerance (total ionizing dose and displacement damage). For gas detectors employing gain stages (e.g., GPDs or GridPix), one must also quantify the spark probability under heavy-ion traversal and verify the robustness of quench/protection circuitry.

Miniaturized HardPix radiation monitors based on the Timepix3 ASIC have already flown to low-Earth orbit (LEO) within the UKRI SWIMMR-1 and SWIMMR-2 programs, thereby assessing Timepix3 operation in space in low-equatorial orbit (LEO) and providing confidence in their SEL/SEU resilience. 
During early GPD development, the detector was irradiated under nominal operating conditions with a 500 MeV nuc$^{-1}$ Fe beam \cite{Bellazzini2010}. The outcome demonstrated that the GPD can tolerate heavy-ion interactions consistent with roughly 20 years of operation in LEO.

While in the next paragraph we report on the expected energy deposit of iron ions in LEO orbit, with IXPE now operational, new X-ray polarimetry missions are being conceived for Sun–Earth Lagrange orbits (like L1 or L2), which offer roughly a factor of two increase in available observing time per source and a more benign thermal environment. Any such mission will still require a tailored radiation/particle environment assessment. Currently, eROSITA aboard the Spectrum-Roentgen-Gamma operates around L2, and Athena is planned for L1. We, therefore, will rescale our calculation of deposited energy for LEO orbit, by considering the measured increase of background in such orbits around the Lagrangian points. The expected background for such unstable gravitational points depends mostly on galactic cosmic rays, the intensity of which is modulated by solar activity and by the related particle wind. We estimate roughly the expected level in L1 or L2 background by comparing its rate in similar devices (e.g., CCDs) flown in LEO and in L1. Such measurements indicate a factor of 5 to 20 increase in background for the eROSITA CCDs compared to the Suzaku CCD\cite{Yamaguchi2006,Perinati2024}, (depending on whether the CCDs are illuminated back or front). We will fold this difference into estimates of ion fluence and SEE rates when assessing total expected ion counts and overall survivability.

\section{PROTON IRRADIATION OF GRIDPIX DETECTORS}

The main contributor to SEE and to the induction of sparks in gas detectors is the energy deposited by ionizing charged particles. This energy deposited in space-based detectors can be evaluated with models that predict the energy loss of moving charged particles that pass through matter. The energy deposition in matter is described by the Bethe-Bloch equation \cite{bethe1932bremsformel},

\begin{equation}\label{eq:BB}
    -\frac{dE}{dx} = \frac{4\pi e^4 z^2}{m_0 v^2}N \times Z \left[ \mathrm{ln}\frac{2m_0v^2}{I} - \mathrm{ln} \left( 1 - \frac{v^2}{c^2} \right) - \left( \frac{v^2}{c^2} \right) \right]
\end{equation}

where, $ze$ is the charge of the particle, $v$ is the velocity of the charged particle, $m_0$ is the electron rest mass, $Z$ is the absorber atomic number, $I$ is the empirical average ionization and excitation potential of the absorber. More energy is deposited by heavier elements (large $z$) that move slowly (small $v$). 

Ionizing radiation in the space environment is classified into two categories. The first category consists of charged particles trapped by the geomagnetic field and is called the Van Allen radiation belt. The Van Allen radiation belt comprises two zones, called the inner and outer belts, which extend from a few hundred to several tens of thousands of kilometers from Earth. The inner belt is dominated by trapped protons (hundreds of MeV), while the outer belt is dominated by trapped electrons (a few MeV). The region of the inner belt closest to Earth's surface, where the proton and electron flux is maximum, is known as the South Atlantic Anomaly (SAA). Satellites in low-Earth orbit (LEO) regularly cross the SAA, during which high-voltage units are temporarily shut down to prevent damage \cite{iwahashi2011heavy}. The second category of ionizing radiation is galactic cosmic rays (GCR). GCR is primarily composed of protons, but it also includes a significant flux of fully ionized heavy atomic nuclei \cite{cronin1999cosmic}. Compared to protons, heavy GCR ions deposit substantially more energy in detectors (Eq~\ref{eq:BB}). Among these heavy ions, iron (Fe) ions are particularly important. Although heavier ions are present, their fluxes are orders of magnitude lower, making Fe the dominant contributor to energy deposition from heavy ions in the detector.

We computed the equivalent radiation dose that the detector would receive over 20 years in LEO, and exposed the detector to that dose on the ground by irradiating it with a proton beam.

\subsection{GCR dose on GridPix}
\label{sec:dose_compute}

\begin{figure}[ht]
\centering
\subfigure[\label{fig:gcr_depo}]{\includegraphics[scale=0.5]{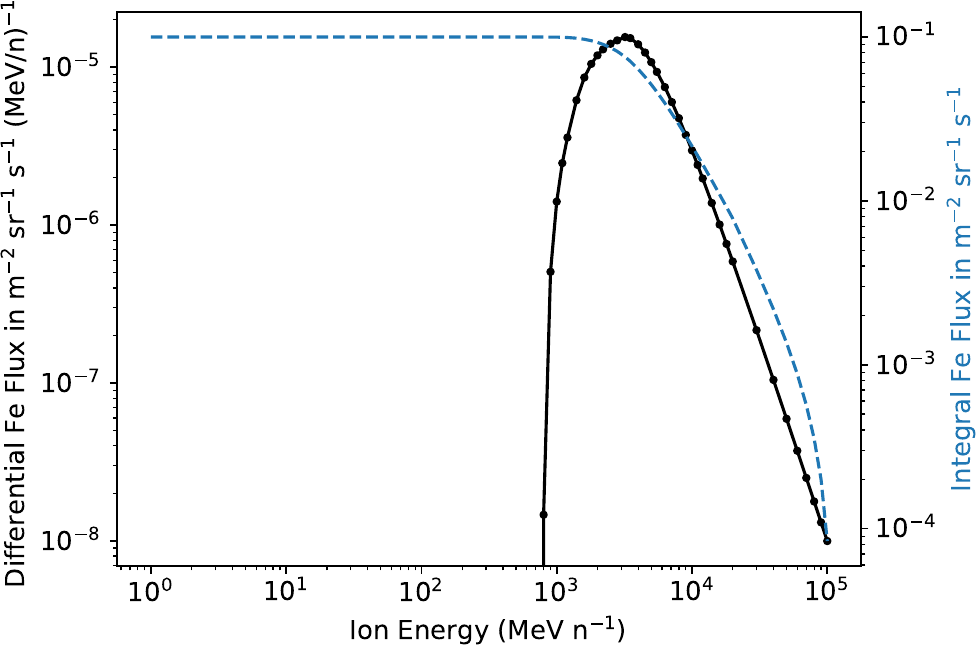}}
\subfigure[\label{fig:detector_depo}]{\includegraphics[scale=0.5]{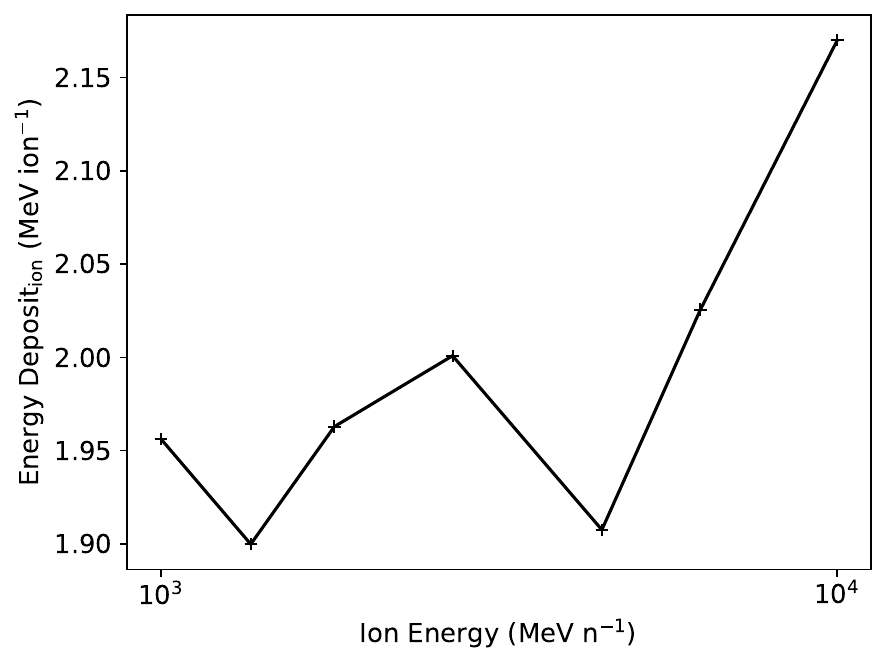}}
\caption{\label{fig:GCR_calculations} Figures show the Fe ion flux and energy deposition in GridPix detector in flight. X-axis in both figures is incident Fe ion energy in MeV nuc$^{-1}$ ({\bf a}) 20 year average differential and cumulative (larger than a given ion energy) GCR flux from Fe ions for 600 km 30$^\circ$ inclined LEO orbit at mission epoch. Geomagnetic field shields ions having energies below $\sim10^3$ MeV nuc$^{-1}$. ({\bf b}) The energy deposited by Fe ions having between $10^3$ and $10^{4}$ MeV nuc$^{-1}$ in the GridPix gas volume after passing through a Beryllium window. Y-axis is the energy deposited by a single Fe ion in MeV.}
\end{figure}

With SPENVIS\footnote{\url{https://www.spenvis.oma.be/}}, we computed the energy spectra of Fe ions from the GCR for a 600 km, 30$^\circ$ LEO for 20 year mission lifetime between 2026 and 2046 (Fig.~\ref{fig:GCR_calculations}). Estimates are based on the ISO-15390 GCR model \cite{iso2004space}. We estimated the energy deposited by these Fe ions in the gas volume of the detector after it passes through a 50 $\mu$m thick Beryllium window. The detector gas volume is assumed to be 1 cm thick, filled with Ar/CO$_2$ in a 70:30 proportion at 1 bar pressure. The density of the gas mixture was set at 1.836 mg cm$^{-3}$. With SRIM-TRIM\footnote{\url{http://www.srim.org/}}, we estimate the energy deposited in the gas mixture by Fe ions with energies between $10^3$ and $10^4$ MeV nuc$^{-1}$.

The integral flux of Fe ions from the GCR above 1000 MeV nuc$^{-1}$ is 9.96$\times10^{-2}$ m$^{-2}$ sr$^{-1}$ s$^{-1}$ (Fig.~\ref{fig:GCR_calculations}). For a detector window area of $10^{-4}$ m$^{2}$ and an opening angle volume of $2\pi$ str, the expected Fe ion rate in orbit greater than 1000 MeV nuc$^{-1}$ (limit of geomagnetic shielding) affecting the detector is $\sim62.5\times10^{-6}$ ions s$^{-1}$. Over 20 years, the total number of Fe ions impinging on the detector is $\sim39,500$ ions. Considering a constant energy deposit of 2.17 MeV per ion (Fig.~\ref{fig:GCR_calculations}), the total energy deposited in the gas cell is estimated to be $\sim86000$ MeV. Our aim is to irradiate the detector with ions so that it deposits at least this much energy in the active gas volume, with the goal of approximately as much as required by a L1 or L2 orbit. 

The experimental setup consisted of a GridPix detector with a circular opening (\diameter8 mm) and a Kapton window on the side wall of the cylindrical housing, allowing the proton beam to pass through the gas mixture into the active volume parallel to the surface of the ASIC and deposit energy. The proton beam traverses a 7.5 mm air gap before entering the gas volume through a 50 $\mu$m thick Kapton window. The tubular housing wall was made of poly(methyl methacrylate) (PMMA), 15mm thick and 30mm high. It was filled with an Ar/CO$_2$ mixture in an 80/20 proportion at 1 bar. 

\begin{figure}[ht]
\centering
\subfigure[\label{fig:render}]{\includegraphics[height=5cm]{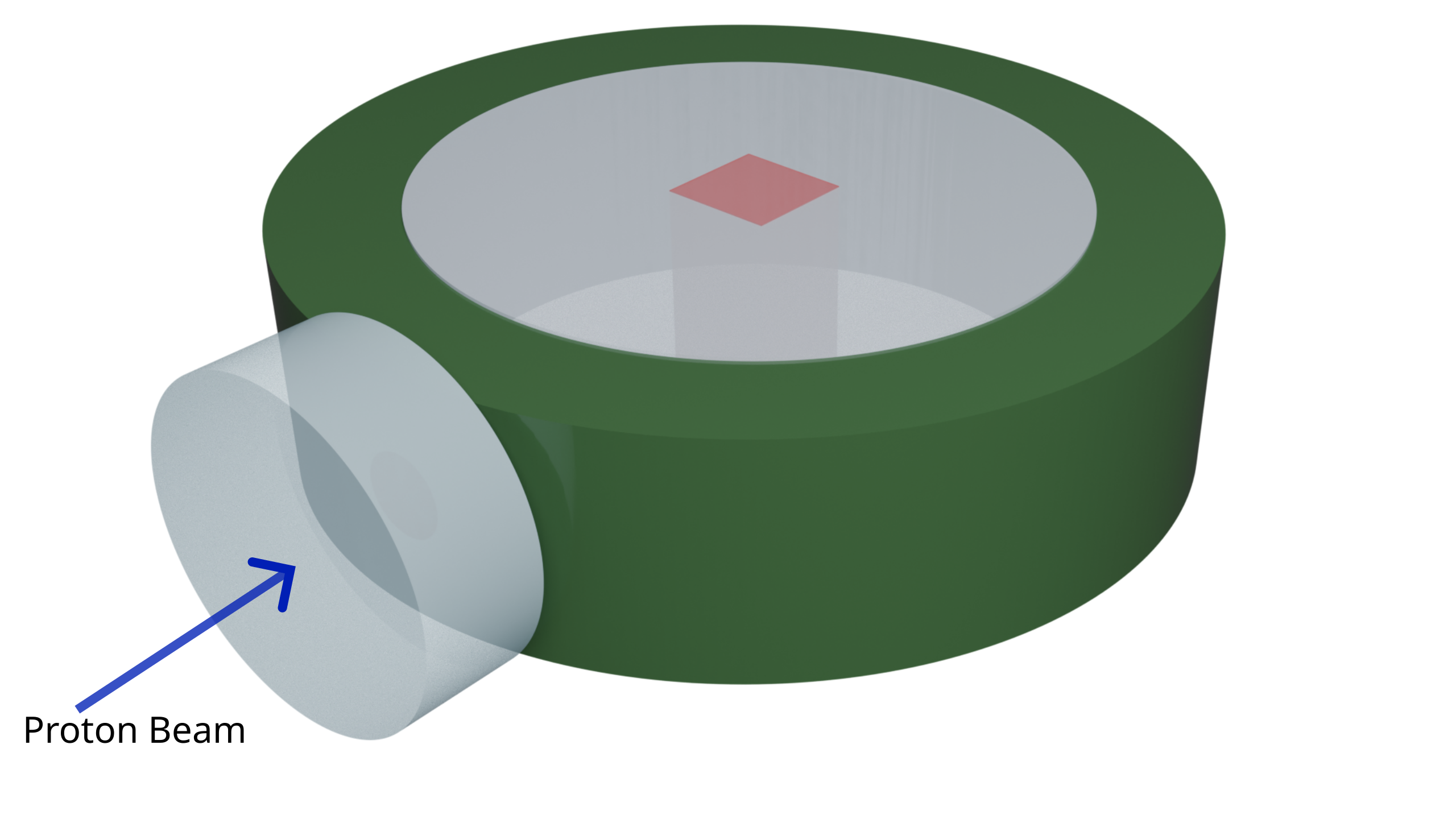}}
\subfigure[\label{fig:edepo}]{\includegraphics[height=6cm]{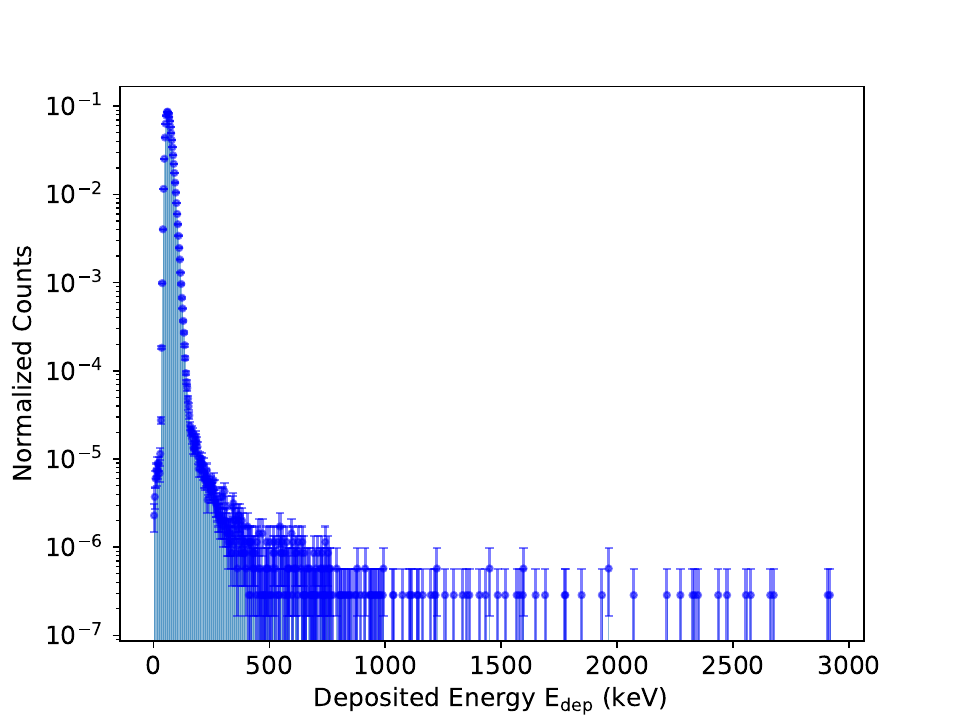}}
\caption{({\bf a}) Rendering of the Geant4 simulation setup used to estimate energy loss by protons in the GridPix detector. The proton beam enters through the air gap (pale blue cylinder, left), passes a Kapton window (brown, visible inside the air gap), and continues into the detector gas volume through an opening in the housing wall. The detector housing wall (green), gas mixture (pale blue), and the ASIC active volume (red) are shown. ({\bf b}) The energy loss spectrum of 13.6 MeV protons in the active volume, the average deposited energy is 64 keV per proton.}
\label{fig:geant4}
\end{figure}

The proton beam passes through the gas volume into the active GridPix volume, defined as a cuboid of $14\times14\times30$ mm$^3$ in the center. We developed a Geant4 simulation framework to estimate the energy deposited by protons in the active gas volume of the aforementioned experimental configuration, as shown in Fig.~\ref{fig:geant4}. The FTFP\_BERT physics list was employed for electromagnetic and hadronic interactions. The hadronic component of this list of physics includes elastic, inelastic, and capture processes, while the electromagnetic component uses standard Geant4 electromagnetic physics \cite{allison2016recent}. 

A 13.6 MeV proton beam with a 2D Gaussian profile (4 mm fwhm) was injected from the air gap into the Kapton window. The energy deposited by protons in the active volume was recorded. The simulation results show that, on average, the proton deposits 64 keV in the active detector volume (Fig.~\ref{fig:geant4}). Therefore, to reproduce the 20-year equivalent space dose of $\sim86$ GeV, a total of $1.3\times10^6$ protons at 13.6 MeV are required.

\subsection{Proton Irradiation}
\label{sec:hiskp}

\begin{figure}[htpb]
\centering
\subfigure[\label{fig:expt_setup}]{\includegraphics[scale=0.22]{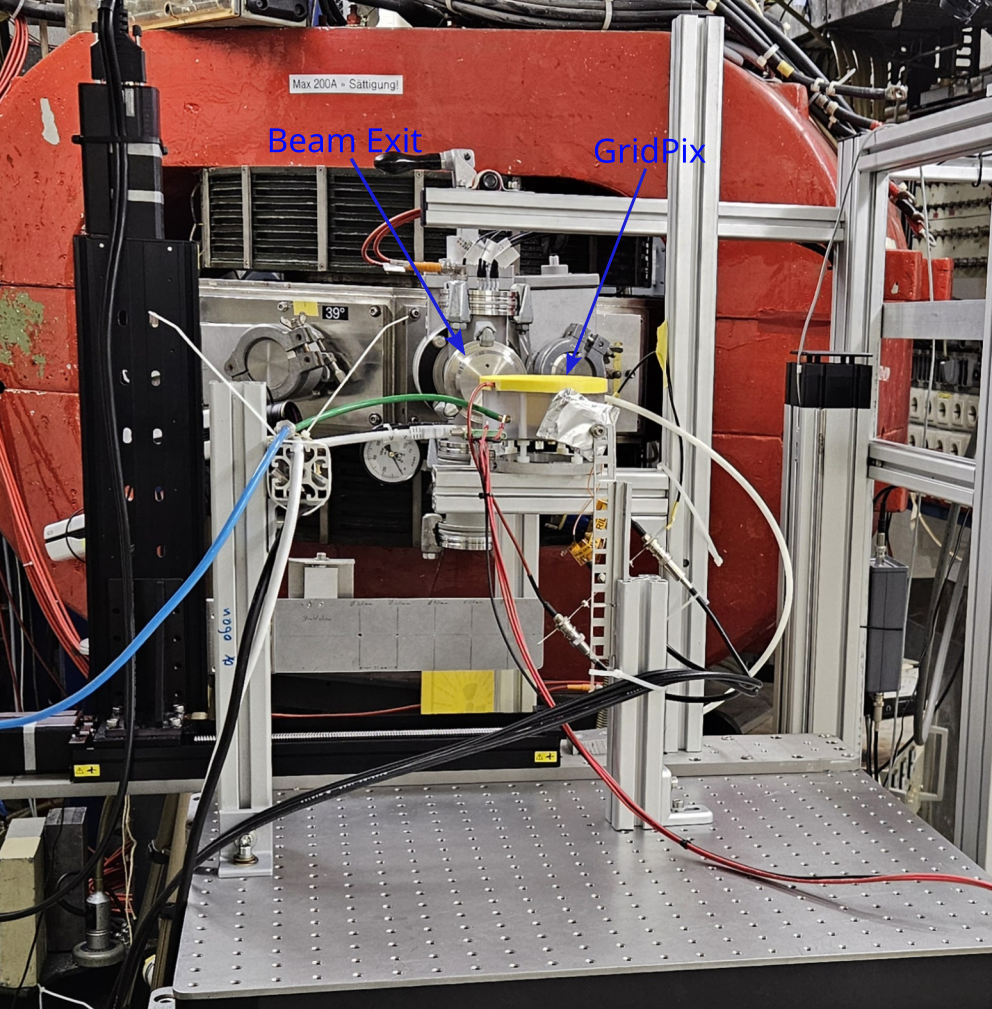}}
\hspace{1cm}
\subfigure[\label{fig:proton_track}]{\includegraphics[scale=0.55]{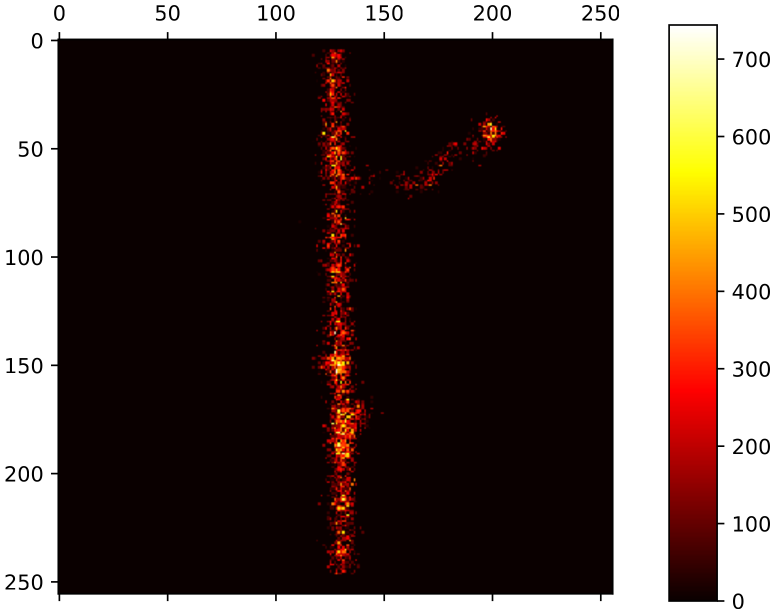}}
\caption{({\bf a}) Experiment setup at the HISKP cyclotron facility. The proton beam was directed into the Gridpix parallel to the ASIC surface through a Kapton window. ({\bf b}) Charge map of a representative proton track recorded by GridPix during irradiation. The X and Y axes correspond to pixel coordinates, while charge deposition is color-coded. The branch emerging from the main track towards the top is a $\delta$-ray.}
\label{fig:Beam_expt}
\end{figure}

Proton irradiation was performed at the Bonn Isochronous Cyclotron facility of the University of Bonn, a facility used for detector testing and nuclear physics research \cite{sauerland2022proton, wolf_cyc, Sauerland:2024nnr} at the Helmholtz Institute for Radiation and Nuclear Physics (HISKP). The accelerator delivers beams of protons, deuterons, or ions up to fivefold ionized nitrogen, with energies between 7 and 14 MeV nuc$^{-1}$ and beam currents up to 1 $\mu$A. For our purpose, we used a 13.512 MeV proton beam (Fig.~\ref{fig:Beam_expt}) at the irradiation site at beamline C.

\begin{figure}
    \centering
    \includegraphics[width=0.4\linewidth]{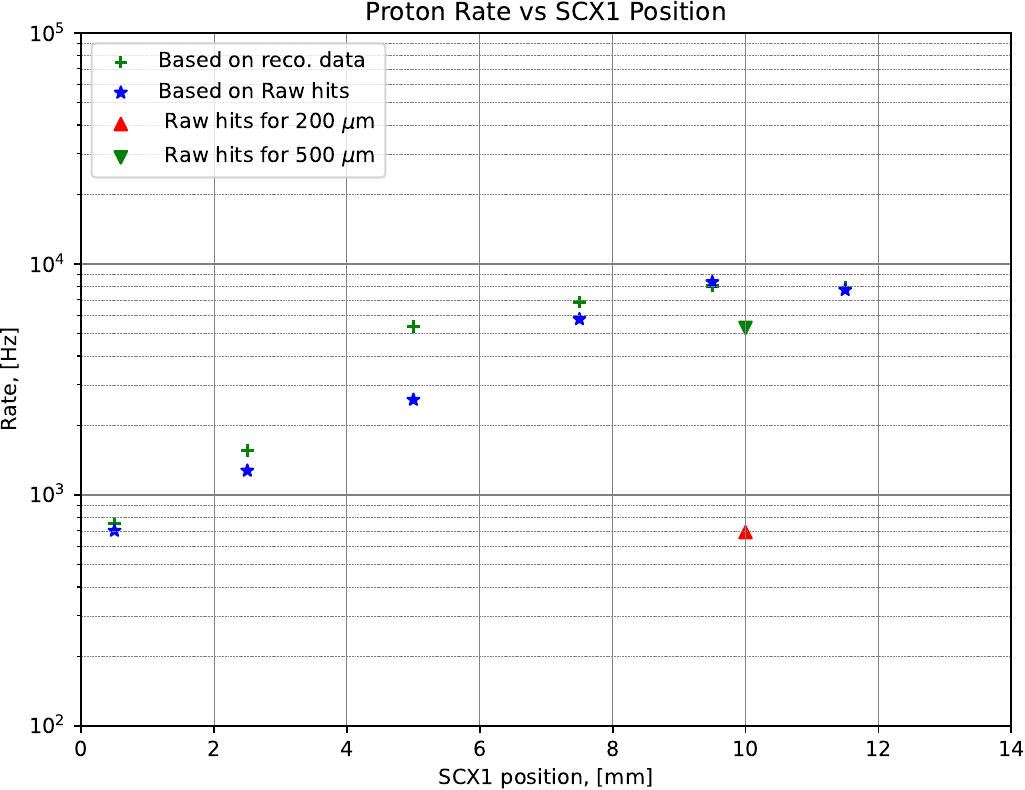}
    \caption{Proton rates measured with the GridPix detector for different aperture openings adjusted with scraper SCX1 position. The measurements were performed while keeping the beam intensity below the operational limits of the beam exit monitor. The observed dependence of the detected rate on aperture size demonstrates stability of the cyclotron beam and its tuning at low intensities.}
    \label{fig:scraper}
\end{figure}

The beam parameters (current and position) were non-destructively monitored by a Secondary Electron Monitor (SEM) at the beamline exit. The beam exhibits a 2D Gaussian profile (standard deviations $\sigma_x\sim3.2$ mm, $\sigma_y\sim4.7$ mm) at the output. The vacuum was maintained in the beam line separating the environment by a 30 $\mu$m thick AlMg$_3$ window. The output beam area could be gradually masked with a horizontal and vertical set of metallic apertures (scrapers), approximately 10 m upstream of the beamline, to adjust the effective current provided to the GridPix. The lowest beam current, detectable by SEM, is on the order of 20 pA. However, our required total irradiation dose was approximately $1.3\times10^6$ protons ($\S$\ref{sec:dose_compute}), while even at 20 pA currents the beam would deliver $100\times$ the required dose within a second. Therefore, the scrapers were closed to a minimum aperture where the beam current was $\ll$ 20 pA. Then they were slightly opened and the proton rate was subsequently measured directly from GridPix (Fig.~\ref{fig:scraper}).

The typical energies of Fe ions from GCR incident on the detector in space are a few $\times10^3$ MeV nuc$^{-1}$ (Fig.~\ref{fig:GCR_calculations}). This raises a question about the instantaneous energy deposited by 13.512 MeV protons in the detector. The deposited energy scales as ${z^2}/{v^2}$ (Eq.~\ref{eq:BB}), where $z$ is the atomic number of the ion and $v$ is the ion velocity. Consequently, a proton with a kinetic energy of 13.512 MeV deposits only a fraction $\sim{{(1^2/13.512)/(26^2/1000)}} \sim0.11\times$ energy of 1000 MeV nuc$^{-1}$ Fe ion instantaneously. Thus, the detector was eventually exposed to a less hostile radiation environment. In fact, the aim of the present test was to demonstrate the use of a cyclotron facility at an affordable low rate for space applications; the final survivability test will be performed by using end-of-range nitrogen ions, with a larger instantaneous energy density deposit. 

The experimental setup consisted of the GridPix detector with a Kapton entrance window on the side walls, aligned with the beam exit diaphragm after an air gap of about 5--10 mm. The detector consisted of a Timepix readout ASIC with an InGrid developed on it. During the irradiation, the detector was operational with InGrid HV on and data acquired in TOT (spectroscopy) mode.

Before irradiating with protons, three kinds of data were acquired: (1) threshold distribution across pixels, before and after threshold equalization, (2) noise scans of each pixel, and (3) spectral acquisition of a radioactive $^{55}$Fe isotope that emits 5.89 keV X-rays. After irradiation, these three procedures were repeated. Any degradation in the detector hardware would manifest itself as changes in these responses. However, no significant deviations were observed relative to the pre-irradiation data. Additionally, no abrupt increase in counts was observed during the run, indicative of no sparking in the InGrid.

From the detector data, the dead-time corrected count rates were calculated to be a total of $157\times10^6$ protons. This corresponds to approximately $100\times$ the required radiation dose expected over 20 years in LEO, but is also larger than that expected in L1 or L2 orbits. However, the instantaneous energy deposition was only a fraction of that of heavy ions. In the next run, we will utilize five-fold ionized nitrogen ions to mitigate this limitation.

\begin{figure}[htpb]
\centering
\subfigure[\label{fig:55Fe_before}]{\includegraphics[scale=0.50]{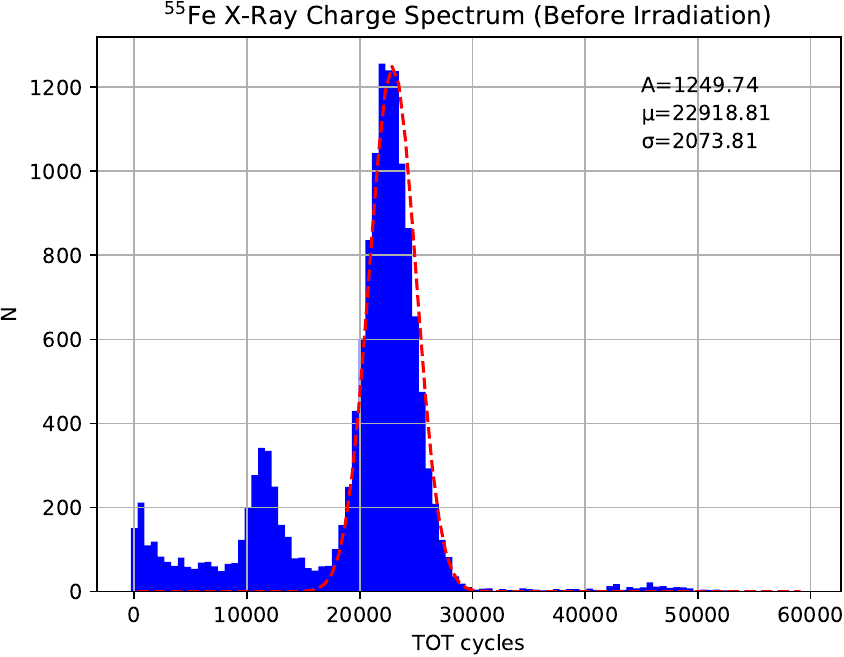}}
\hspace{1cm}
\subfigure[\label{fig:55Fe_after}]{\includegraphics[scale=0.50]{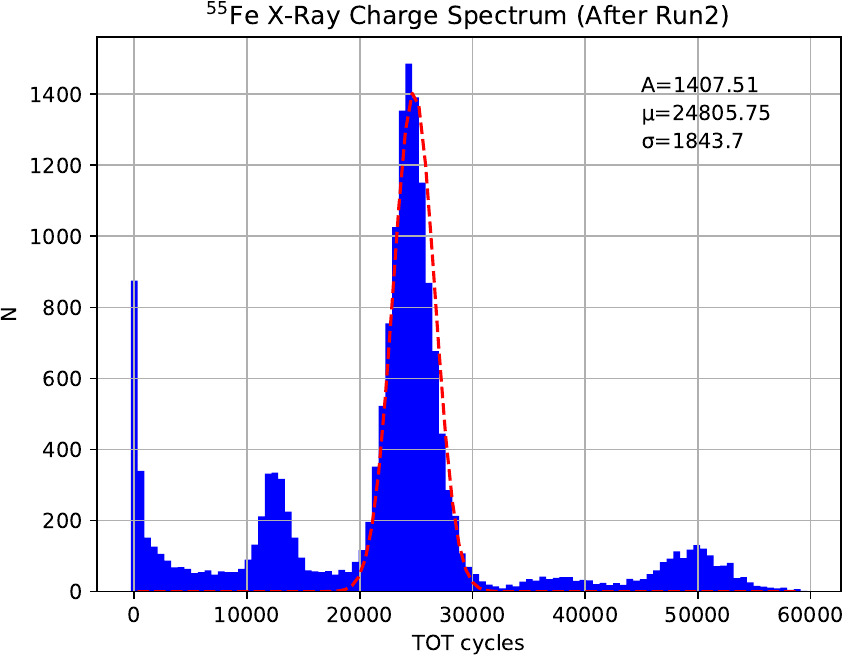}}
\caption{X-ray spectrum of $^{55}$Fe (5.89 keV) recorded with the GridPix detector ({\bf a}) before proton irradiation and ({\bf b}) after proton irradiation. The detector response remains stable, with gain and energy resolution staying within acceptable limits. A shift in peak position indicates a gain change of approximately 8\%, and the energy resolution changed from 21\% to 17\%. This gain shift between the two measurements may be due to a change in the environmental temperature after the second irradiation. A slight increase in counts in the first energy channel is visible after the second irradiation; this feature is under investigation. These additional counts cannot be attributed to a substantial increase in electronic noise, given the improved energy resolution of the spectrum shown in panel \textbf{b}. The peaks to the right of the main 5.89 keV line are attributable to double-photon events, where two photons were detected within the same frame by GridPix. Because the source position differed between the two measurements, the relative prominence of these peaks also differs.}
\label{fig:fe55r}
\end{figure}

\section{Conclusions}
In this work, we report preliminary activities related to space radiation tolerance tests of the GridPix detectors. The expected radiation dose for a 20-year mission in LEO was estimated, and the corresponding proton dose in the detector's active volume was determined using Geant4 simulations. The detector was then irradiated with a proton beam directed towards the active region and parallel to the active surface of the read-out ASIC at the Bonn Isochronous Cyclotron. The cyclotron could be successfully operated at sufficiently low proton rates for our study. The results demonstrate that the detector maintains stable performance even after exposure to significant proton doses, paving the way for future tests with heavier ions. In fact, in the next run, we plan to irradiate the GridPix detector having a Timepix readout ASIC with fivefold ionized nitrogen ions (14 N5+), which will deposit more instantaneous energy than the space environment,  with the beam directed into the readout ASIC. The scope of this test is to verify the survivability of the system ASIC+Ingrid to possible spark induced by the nitrogen ion.

\acknowledgments 
 We acknowledge the partial contribution of Progetti di Ricerca di Rilevante Interesse Nazionale del Ministero
della Università e della ricerca scientifica (PRIN-MUR) HypeX: High Yield Polarimetry Experiment in X-rays
(Hype-X) prot. 2020MZ884C Parts of this work have received funding from the German Federal Ministry of
Education and Research under grant no. 05K22PD1.

\bibliography{report} 
\bibliographystyle{spiebib} 

\end{document}